\documentclass[useAMS,usenatbib]{mn2e}
\usepackage{graphics,graphicx}
\usepackage{amssymb}
\usepackage{amsmath} 
\usepackage{latexsym} 

\title[CN excitation and electron densities in diffuse molecular
  clouds]{CN excitation and electron densities in diffuse molecular
  clouds} \author[Stephen Harrison, Alexandre Faure and Jonathan
  Tennyson]{Stephen Harrison$^1$\thanks{E-mail:
    stephen.harrison@ucl.ac.uk}, Alexandre Faure$^2$\thanks{E-mail:
    afaure@obs.ujf-grenoble.fr} and Jonathan
  Tennyson$^1$\thanks{E-mail: j.tennyson@ucl.ac.uk}\\ $^1$ Department
  of Physics and Astronomy, University College, London, Gower St.,
  London WC1E 6BT, UK\\ $^2$ UJF-Grenoble 1 / CNRS-INSU, Institut de
  Plan\'etologie et d'Astrophysique de Grenoble (IPAG) UMR 5274,
  Grenoble, F-38041, France}

\begin{document}
\date{Accepted ? Received ?}

\maketitle

\begin{abstract}
Utilising previous work by the authors on the spin-coupled
rotational cross-sections for electron-CN collisions, 
data for the associated rate coefficients is presented. 
Data on rotational, fine-structure and hyperfine-structure transition
involving
rotational levels up to  $N$=20 are computed for temperatures in the range
10 -- 1000~K. Rates are calculated by combining Born-corrected R-matrix
calculations with the infinite-order-sudden (IOS) approximation.
The dominant hyperfine transitions are those with $\Delta
N=\Delta j= \Delta F=1$. For dipole-allowed transitions,
electron-impact rates are shown to exceed those for excitation of CN
by para-H$_2$($j=0$) by five orders of magnitude. The role
of electron collisions in the excitation of CN
in diffuse clouds, where local excitation competes with the cosmic
microwave background (CMB) photons, is considered. Radiative transfer
calculations are performed and the results compared to observations. 
These comparisons suggest that 
electron density lies in the range $n(e)\sim 0.01-0.06$~cm$^{-3}$ for
typical physical conditions present in diffuse clouds.
\end{abstract}


\section{Introduction}

Soon after its discovery by \cite{penzias65}, the cosmic microwave
background (CMB) was postulated as primarily responsible for the
rotational excitation of CN observed in diffuse clouds
\citep{thaddeus66,field66}. Optical absorption-line measurements of
interstellar CN have thus long been used to estimate the temperature
of CMB radiation at 2.6 and 1.3~mm, the wavelengths of the two lowest
CN rotational transitions \citep{thaddeus72}. It was soon realized,
however, that the accuracy of this indirect method is limited by line
saturation and local collisional excitation effects. Since the first
high accuracy measurements of the CMB temperature by the {\it COBE}
satellite \citep{mather90}, with the latest value at $T_{\rm
  CMB}=2.72548\pm 0.00057$~K \citep{fixsen09}, CN absorption line
observations have been used to provide an independent calibration of
the {\it COBE} satellite, to sample the CMB far from the near-Earth
environment, and to measure the rotational excitation of CN in excess
of $T_{\rm CMB}$, i.e. the local excitation processes. Differences
between the {\it COBE} results and those from CN have recently
been discussed by \citet{Leach12}.

CN absorption lines with very high signal-to-noise ratio were observed
recently by \cite{ritchey11} along 13 lines of sight through diffuse
molecular clouds. Their careful analysis of the CN rotational
excitation implies a mean excess over the temperature of the CMB of
only 29$\pm 3$~mK, which is significantly lower than previous
measurements. If electron-impact is the dominant local CN excitation
process, as it is generally assumed, then the excess temperature can
yield an estimate of the electron density in the gas
\citep{BV1991}. The electron density is a crucial parameter for
modelling both the physics and chemistry of molecular clouds. It is
generally estimated from the observation of ultraviolet lines of
atomic species like C and C$^+$. In clouds of modest density ($n({\rm
  H_2})\lesssim$ 1000~cm$^{-3}$) the fractional ionization
($x_e=n(e)/n({\rm H_2}$) is thus typically 10$^{-5}$ -- 10$^{-4}$. An
accurate and independent determination of the electron density from CN
excitation obviously requires a good knowledge of the electron-impact
excitation rate coefficients.

The first cross section calculations for the electron-impact
rotational excitation of CN were based on the Born approximation
\citep{thaddeus66}. More accurate close-coupling calculations were
then performed \citep{AD1971} and these were found to agree with
the Born results (for the $N=0\rightarrow 1$ transition) within a
factor of 3 above $\sim$15~K \citep{thaddeus72}. More recently, we
have revisited the rotational excitation of the CN radical using the
R-matrix approach combined with the infinite-order-sudden (IOS)
approximation to derive, for the first time, electron-impact
spin-coupled cross sections \citep{jt531}. Our calculations were
restricted to electron energies above 0.1~eV and the high energy
results were found to be heavily influenced by both the A~$^2\Pi$ and
B~$^2\Sigma^+$ excitation thresholds at 1.52 and 3.49~eV,
respectively. At energy below these thresholds, however, the usual
propensity rule for parity-conserving transitions ($\Delta j=\Delta
N$) was found to hold.

In this work, we extend the calculations of \citet{jt531} to lower
collision energies in order to derive rate coefficients down to the
low temperatures of the interstellar medium. In addition to the
spin-doubling of CN, we consider also the hyperfine
structure. In Section~2, the method employed to derive the fine and
hyperfine rate coefficients is outlined and comparisons with other
sets of collisional data are presented. In Section~3, the results of
radiative transfer calculations, including the CMB radiation and local
excitation caused by electron and neutral collisions, are presented
and compared to observational results. Conclusions are drawn in
Section~4.

\section{Rate coefficient calculations}

Electron-impact rate coefficients for both fine and hyperfine
transitions were calculated from the pure rotational rate coefficients
using the IOS formalism. The rotational cross sections were computed
as in \cite{jt531} by combining R-matrix calculations \citep{jt474, jt514}, Born corrected
for dipolar transitions \citep{np82}, with the adiabatic-nuclei-rotation (ANR)
approximation, which is very similar to the infinite-order sudden (IOS) approximation. Both
approximations consist of assuming that the target rotational states
are degenerate, which is valid when the rotational spacings are
negligible with respect to collisional energy. In practice, cross
sections were obtained for collision energies above 10~meV and they
were corrected using a kinematic ratio to account for the rotational
spacings, as in \cite{jt531}. They were finally extrapolated down to
the rotational thresholds using the procedure described in
\cite{rabadan98}, see eq.~(1) of their paper, which was calibrated
using the rotational close-coupling results of \cite{AD1971}. Assuming
that the electron velocity distribution is Maxwellian, rate
coefficients were obtained for temperatures in the range 10 –- 1000~K and
for transitions among all levels up to $N$ = 20. A similar study by \citet{jt414}
considered hyperfine structure in electron collisions with the electron spin singlet
HCN/HNC system which therefore does not display fine structure splitting. 

Within the ANR or IOS formalism, the spin-coupled or fine structure
rate coefficients (and cross sections) can be obtained from the
fundamental pure rotational cross sections, i.e. those out of the
lowest $N=0$ level, as follows (see \cite{jt531} and references
therein):
\begin{eqnarray}
\label{fine}
& & k^{IOS}_{Nj \to N'j'} (T)  =  (2N+1)(2N'+1) (2j'+1)\sum_{\lambda} \nonumber \\
& & \times \left( 
\begin{array}{ccc}
N' & N & \lambda \\
0 & 0 & 0 
\end{array}
\right)^2
\left\{
\begin{array}{ccc}
\lambda & j & j' \\
S & N' & N 
\end{array}\right\}^2 \times  k_{0 \to \lambda}(T),
\end{eqnarray}
where $N$ is the rotational angular momentum of CN, $S$ is the
electron spin (here $S=$1/2), $\textbf{j}=\textbf{N}+\textbf{S}$ and
$k_{0 \to \lambda}(T)$ is the pure rotational rate coefficients out of
the lowest $N=0$ level. In practice, as the rotational cross sections
were corrected for threshold effects and extrapolated, Eq.~\ref{fine}
is expected to be accurate only above $T\sim 100$~K. We have therefore
implemented the \lq scaling\rq\ method originally proposed by
\cite{neufeld94} in which the spin-coupled rate coefficients are
obtained as:
\begin{equation}
\label{finescal}
  k_{Nj \to N'j'} (T) = \frac{k^{IOS}_{Nj \to N'j'}(T)}{k^{IOS}_{N\to
    N'}(T)}k_{N\to N'}(T),
\end{equation}
where 
\begin{equation}
\label{iosrot}
k^{IOS}_{N \to N'}(T)=(2N'+1)\sum_\lambda \left(\begin{array}{ccc}
N' & N & \lambda \\ 
0 & 0 & 0
\end{array}\right)^{2} k_{0 \to \lambda}(T).
\end{equation}
The scaling of Eq.~(\ref{finescal}) guarantees in particular the following
equality:
\begin{equation}
\label{sum}
\sum_{j'} k_{Nj \to N'j'}(T)=k_{N\to N'}(T).
\end{equation}
We also note that in Eqs.~(\ref{fine}) \& (\ref{iosrot}) the fundamental
excitation rates $k_{0 \to \lambda}$ were replaced by the corresponding
de-excitation rates using the detailed balance relation, as suggested
by \cite{faure12}:
\begin{equation}
\label{deex}
k_{0\to \lambda}(T)=(2\lambda+1)k_{\lambda \to 0}(T).
\end{equation}

Similarly, the rate coefficients among hyperfine structure levels $(N,
j, F)$ can be obtained from the fundamental spin-coupled rate
coefficients $k_{0,1/2 \to L,L+1/2}(T)$ using the following formula
\citep{faure12}:
\begin{eqnarray}
\label{ioshf}
& & k^{IOS}_{NjF \to N'j'F'} (T)  =  (2j+1)(2j'+1) (2F'+1)\sum_{\lambda} \frac{2\lambda+1}{\lambda+1}  \nonumber \\
& & \times \left( 
\begin{array}{ccc}
j' & \lambda & j \\
-1/2 & 0 & 1/2 
\end{array}
\right)^2
\left\{
\begin{array}{ccc}
j & j' & \lambda \\
F' & F & I 
\end{array}
\right\}^2 \nonumber \\ & & \times \frac{1}{2}[1-\epsilon
  (-1)^{j+j'+L}] k_{0,1/2 \to L,L+1/2}(T).
\end{eqnarray}
where $\epsilon$ is equal to $+1$ if the parity of initial and final
rotational $Nj$ level is the same or $-1$ if the parity of initial and
final rotational $Nj$ level differ. As above, a similar scaling was
implemented:
\begin{equation}
\label{hfscal}
  k_{NjF \to N'j'F'} (T) = \frac{k^{IOS}_{NjF \to N'j'F'}(T)}{k^{IOS}_{Nj\to
    Nj'}(T)}k_{Nj\to Nj'}(T),
\end{equation}
where
\begin{eqnarray}
\label{iosfin}
&  & k^{IOS}_{Nj \to N'j'}(T)  =  (2j'+1)\sum_\lambda \frac{2\lambda+1}{\lambda+1} \left(\begin{array}{ccc}
j' & \lambda & j \\ 
-1/2 & 0 & 1/2
\end{array}\right)^{2} \nonumber \\ 
& & \hspace{1cm} \times \frac{1}{2}[1-\epsilon (-1)^{j+j'+\lambda}]
k^{IOS}_{0,1/2 \to \lambda,\lambda+1/2}(T).
\end{eqnarray}
Finally the fundamental excitation rates $k_{0, 1/2 \to \lambda,
  \lambda+1/2}$ were replaced by the corresponding de-excitation rates
using detailed balance. 

Full details on the above procedure can be found in \cite{faure12}
where scaled IOS rate coefficients are compared in detail with
almost exact recoupling calculations on CN-H$_2$. The scaled IOS
method was found by these authors to reproduce the recoupling results
within a factor of 3 or better, down to very low temperature. Results
should be even better for electron collisions since the electron
motion is much more rapid than H$_2$ and the adiabatic rotational
approximation holds at lower temperature. It should be also
emphasized that the IOS method properly includes the recoupling
algebra, via the 3-$j$ and 6-$j$ coefficients, and the propensity
rules $\Delta j=\Delta N$ (parity-conserving) and $\Delta j=\Delta F$
are correctly predicted. \cite{faure12} showed that these rules
play an important role in radiative transfer applications when line
saturation is important.

\begin{figure}
\begin{center}
	\includegraphics*[scale=0.35]{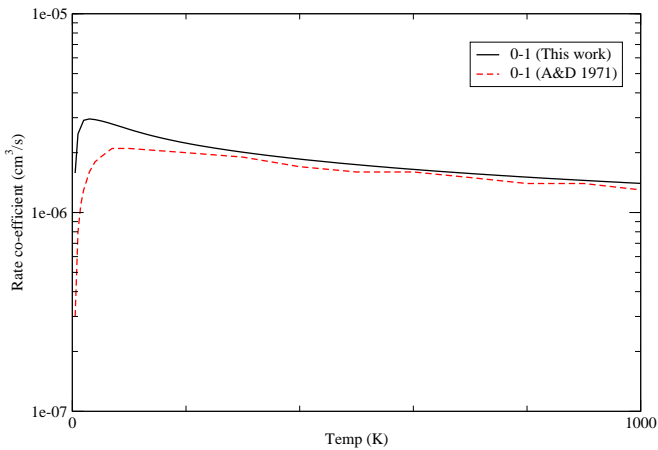}
\end{center}
	\caption{A comparison of the $0\to 1$ rotational rate of this
          work with the rate of \citet{AD1971}.}
	\label{fig:00-01.eps}
\end{figure}

\begin{figure}
\begin{center}
	\includegraphics*[scale=0.35]{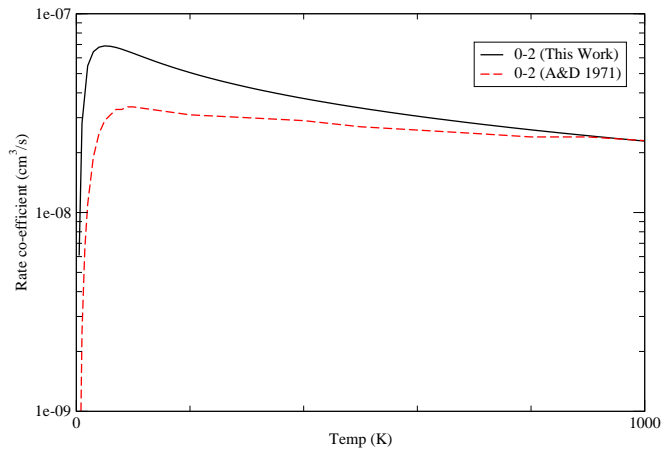}
\end{center}
	\caption{A comparison of the $0\to 2$ rotational rate of this
          work with the rate of \citet{AD1971}.}
		\label{fig:00-02.eps}
\end{figure}

Figures \ref{fig:00-01.eps} \& \ref{fig:00-02.eps} present the
rotational rates $0\to 1$ and $0\to 2$ as a function of temperature,
up to 1000~K, including a comparison with the relevant data from the
work of \cite{AD1971} which has been so far employed in the
astronomical literature. It is clear that the present rotational rates
are larger than those of \cite{AD1971}, particularly at temperatures
below 100~K where our data is about a factor of 2 larger at the
peaks. These differences reflect both the short-range treatment of the
interaction and the extrapolation  at very low energy.

\begin{figure}
\begin{center}
	\includegraphics*[scale=0.45]{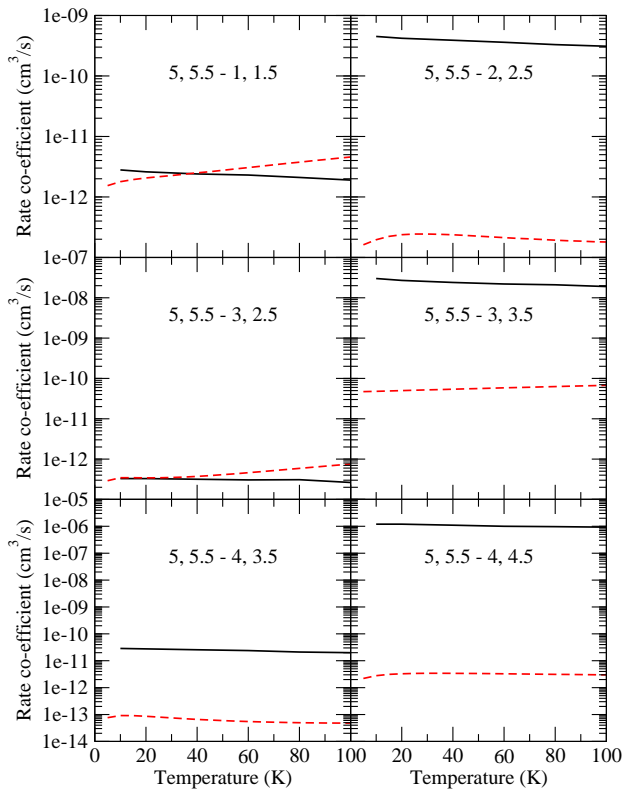}
\end{center}
	\caption{A comparison of the fine structure rate from the $(N,
          j)= (5, 5.5)$ initial level between this work (solid line)
          and the rate of \citet{KLK2012} (dashed line).}
		\label{fig:fs.eps}
\end{figure}

Figure \ref{fig:fs.eps} shows the fine structure e-CN collision rates
out of the $N$=5, $j$=5.5 initial level in comparison with the
relevant data from the work of \cite{KLK2012} for CN colliding with
para-H$_2$($j=0$). We can notice that dipolar transitions with $\Delta
N=1$ have the largest rates for e-CN, in contrast to CN-H$_2$($j=0$)
collisions where transitions with $\Delta N=2$ are preferred. We note,
however, that dipolar transitions are also favoured in the case of CN
colliding with rotationally excited H$_2$ ($j>0$) (Lique, private
communication). For both systems, the propensity rule $\Delta j=\Delta
N$ (i.e. parity-conserving transitions) is observed. As a result, the
favoured transitions are $(N, J)=(5, 5.5)\to (4, 4.5)$ and $(5, 5.5)\to
(3, 3.5)$ for electron and H$_2$($j=0$) collisions, respectively, and
they differ by about 4 orders of magnitude. Finally, we note that the
temperature dependences are very weak (for these de-excitation
transition) in the 10-100~K range.

\begin{figure}
\begin{center}
	\includegraphics*[scale=0.5]{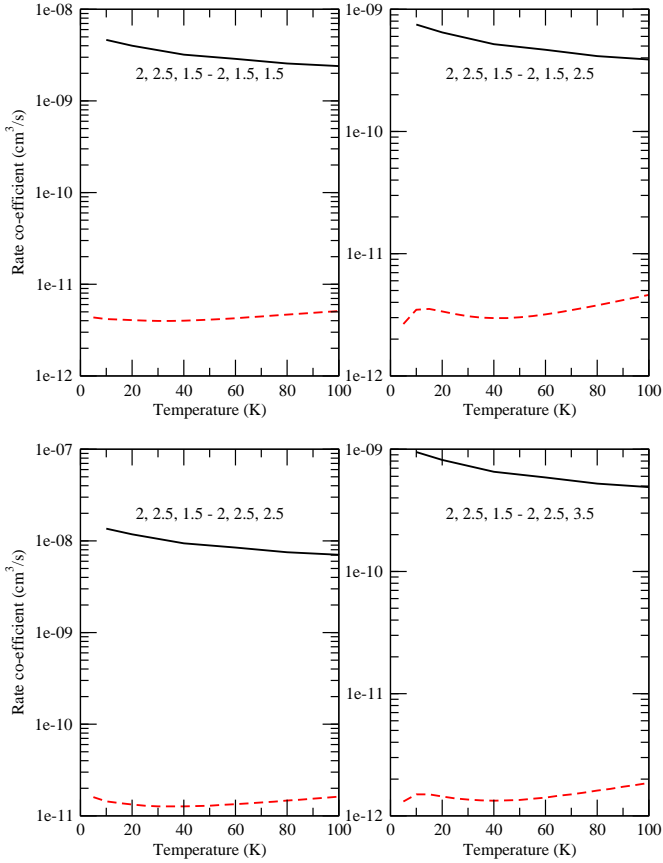}
\end{center}
	\caption{A comparison of the hyperfine structure rate from the
          $(N, J, F)=(2, 2.5, 1.5)$ initial level between this work
          (solid line) and the rate of \citet{KLK2012} (dashed line).}
	\label{fig:hfs_2_2.5_1.5.eps}
\end{figure}

\begin{figure}
\begin{center}
	\includegraphics*[scale=0.5]{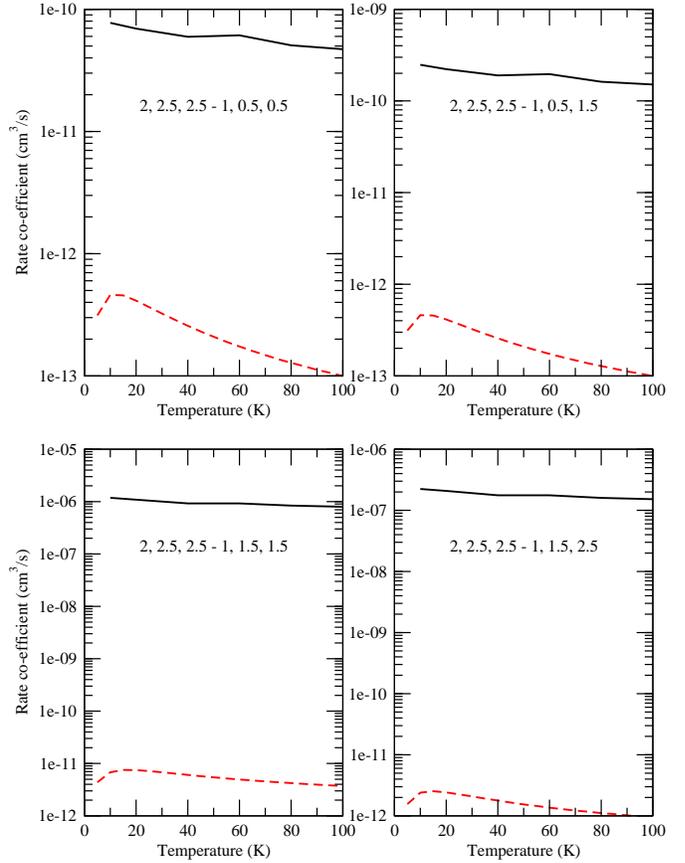}
\end{center}
	\caption{A comparison of the hyperfine structure rate from the
          $(N, J, F)=(2, 2.5, 2.5)$ initial level between this work
          (solid line) and the rate of \citet{KLK2012} (dashed line).}
		\label{fig:hfs_2_2.5_2.5.eps}
\end{figure}

\begin{figure}
\begin{center}
	\includegraphics*[scale=0.5]{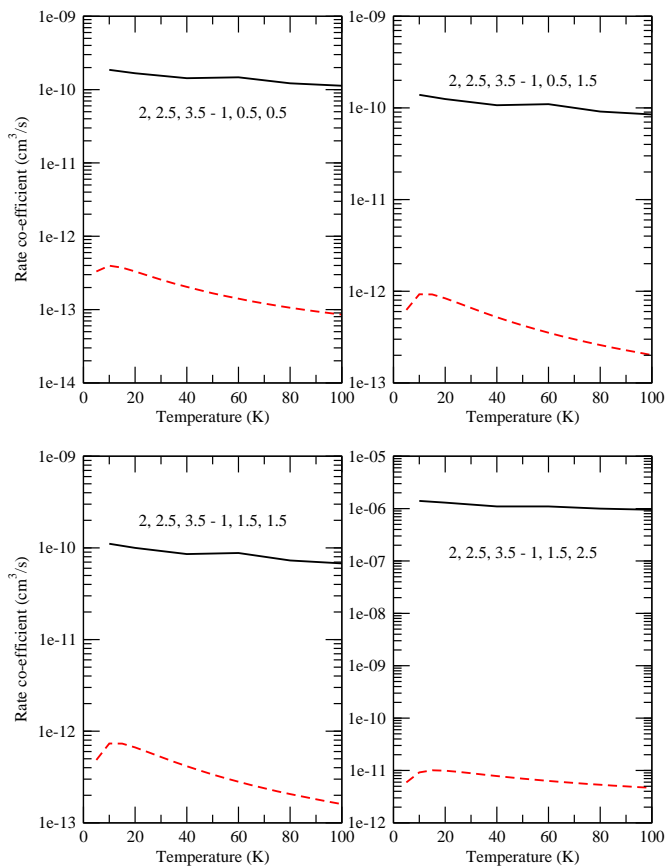}
\end{center}
	\caption{A comparison of the hyperfine structure rate from the
          $(N, J, F)=(2, 2.5, 3.5)$ initial level between this work
          (solid line) and the rate of \citet{KLK2012} (dashed line).}
		\label{fig:hfs_2_2.5_3.5.eps}
\end{figure}

Figures \ref{fig:hfs_2_2.5_1.5.eps}, \ref{fig:hfs_2_2.5_2.5.eps}
\& \ref{fig:hfs_2_2.5_3.5.eps} also give the hyperfine structure rate
comparisons between this work and \cite{KLK2012} for the transitions
out of the $N$ = 2, $j$ = 2.5, $F$ = (1.5 - 3.5) respectively. As
expected, the highest electron-impact rate is observed for the dipolar
transitions $(2, 2.5, 2.5)\to (1, 1.5, 1.5)$ and $(2, 2.5, 3.5)\to (1,
1.5, 2.5)$ corresponding to $\Delta N= \Delta j= \Delta F=1$. These
rates are about 5 orders of magnitude larger than the corresponding
rates for H$_2$($j=0$). For other transitions, the differences range
between 2 and 5 orders of magnitude.

We conclude that the electron-impact excitation of CN should be
significant as soon as the electron fraction $x_e=n({\rm e})/n({\rm
  H_2})$ exceeds $\sim$ 10$^{-5}$ and that these collisions will
strongly favour transitions with $\Delta N= \Delta j= \Delta F=1$, in
contrast to H$_2$($j=0$) collisions which favour $\Delta N= \Delta j=
\Delta F=2$. The present data will be made available in the BASECOL
database \citep{jt547}.

\section{Radiative transfer calculations}

As explained in the Introduction, in diffuse molecular clouds the
rotational excitation of CN is controlled by a competition between the
collisional excitation and the interaction with the CMB
radiation. This competition results in an excess of the CN rotational
excitation over the temperature of the CMB (2.725~K). The CN
excitation temperature, determined from optical absorption lines, is
thus defined as:
\begin{equation}
T_{ex}({\rm CN})=T_{\rm CMB}+T_{\rm loc},
\end{equation}
where $T_{\rm loc}$ is the contribution due to local excitation
mechanism and $T_{ex}$ is determined through the Boltzmann equation:
\begin{equation}
\frac{N(i)}{N(j)}=\frac{g_i}{g_j}\exp(-\frac{h\nu_{ij}}{k_BT_{ex}}),
\end{equation}
where $N(i)$ and $N(j)$ are the column densities of the upper and
lower rotational states, respectively, and $g(i)$ and $g(j)$ are the
corresponding statistical weights. In practice, only the 3 lowest
rotational states are significantly populated, i.e. $N$=0, 1 and 2,
yielding the measured excitation temperatures $T_{01}$(CN) and
$T_{12}$(CN). The local excitation effects can be also directly determined
from a measurement of CN millimetre emission which is unfortunately
weak and rarely detected.

Observationally, the most recent CN optical absorption-line
measurements have provided a weighted mean value of $T_{01}({\rm
  CN})=2.754\pm 0.002$~K, implying an excess over the temperature of
the CMB of $T_{\rm loc}=29\pm 3$~mK \citep{ritchey11}. We note that the
dispersion of these measurements is quite large, i.e. 134~mK, with
some sight lines showing (unphysical) excitation temperature below
$T_{\rm CMB}$. It is generally assumed that electron-impact excitation
is the dominant contribution to this excess temperature. Below we
investigate the influence of varying the electron density on the local
excitation, using a radiative transfer code combined with the best
available electron and neutral collisional rates: those from this work
for electrons and those of \cite{KLK2012} for para-H$_2$($j=0$),
assumed to be identical for hydrogen atoms\footnote{We note that the
  rate coefficients for ortho-H$_2$($j=1$) exceed those for
  para-H$_2$($j=0$) by up to a factor of 10 (Lique, private
  communication). However ortho-H$_2$($j=1$) can be neglected here
  since its abundance in cold ($T<30$~K) diffuse clouds is expected to
  be at least 30 times lower than that of para-H$_2$($j=0$).}. This
kind of analysis has been previously performed by \cite{BV1991} with
old collision data.

\begin{figure}
\begin{center}
\rotatebox{-90}{\includegraphics*[scale=0.4]{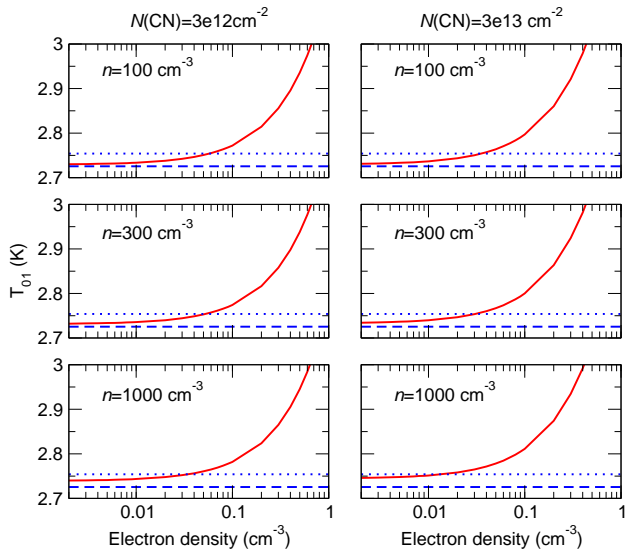}}
\end{center}
	\caption{Excitation temperature $T_{01}$(CN) as a function of
          electron density for different densities ($n=n_{\rm
            H}+n_{\rm H_2}$) and CN column densities, at a single
          kinetic temperature of 20~K. Here the dashed line represents
          the CMB at 2.725~K while the dotted blue line gives the
          measured average excitation temperature at 2.754~K
          \citep{ritchey11}.}
	\label{fig:cn.eps}
\end{figure}

Radiative transfer calculations were performed with the \texttt{RADEX}
code \citep{vandertak07}, using the Large Velocity Gradient (LVG)
approximation for an expanding sphere. The kinetic temperature was
fixed at $T$=20~K, as in the calculations of \cite{ritchey11}. The
line width (FWHM) was taken to be 1.0~kms$^{-1}$, corresponding to a
Doppler line broadening parameter $b$ of 0.6 kms$^{-1}$. The column
density was taken at two typical values $N({\rm CN})=3\times 10^{12}$
and $3\times 10^{13}$~cm$^{-2}$. The density of neutral collision
partners ($n=n({\rm H})+n({\rm H_2})$) was fixed at three
representative values: 100, 300 and 1000~cm$^{-3}$. Finally the
electron abundance was varied from $2\times 10^{-3}$ to 1~cm$^{-3}$,
corresponding to electron fractions $n(e)/n$ in the range $2\times
10^{-6}$ to $10^{-2}$. Results are presented in
Fig.~\ref{fig:cn.eps}. In each panel, the excitation temperature
$T_{01}$(CN) is plotted as a function of the electron abundance. It
should be noted that our excitation calculations provide the
populations of hyperfine levels, from which $T_{01}$ was computed by
summing over hyperfine sublevels. The dashed horizontal line gives the
CMB at 2.725~K while the horizontal dotted line gives the measured
excitation temperature $T_{01}$ at 2.754~K. We first observe that the
excess temperature of 29~mK cannot be reproduced at very low electron
density, indicating that neutral collisions alone cannot explain the
local excitation of CN. This confirms the conclusions of past
investigators \citep{thaddeus72,BV1991}. Second, it can be noticed
that the local excitation is reproduced for a rather restricted range
of electron densities: from 0.01~cm$^{-3}$ at $n$=1000~cm$^{-3}$ to
0.06~cm$^{-3}$ at $n$=100~cm$^{-3}$ with a weak dependence on the CN
column density. Assuming hydrogen is entirely molecular, these
electron density correspond to electron fractions ($x_e=n(e)/n({\rm
  H_2})$) in the range $10^{-5}-6\times 10^{-4}$. This is consistent with
the abundance of interstellar C$^{+}$ ($n({\rm C}^+)/n({\rm H_2})\sim
3\times 10^{-4}$) which is the main source of electrons in the diffuse
interstellar medium.

\begin{figure}
\begin{center}
\rotatebox{-90}{\includegraphics*[scale=0.4]{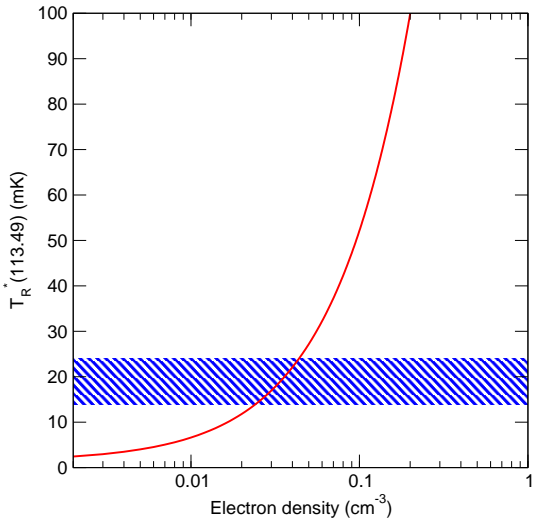}}
\end{center}
	\caption{Plot of the intensity of the line at 113.49~GHz as a
          function of electron density for $T$=20~K, $n$=150~cm$^{-3}$
          and $N$(CN)=2.7$\times 10^{13}$~cm$^{-2}$. Here the blue
          hatched zone shows the observed antenna temperature.}
	\label{fig:tmb.eps}
\end{figure}

In fact, more accurate determination of the electron density can be
achieved for clouds where the physical conditions are reasonably well
known. For instance, the kinetic temperature and the collision density
were determined for the source HD~154368 from the analysis of C$_2$
excitation by \cite{sonnentrucker07}. These authors found
$T=20\pm 5$~K and $n=150^{+50}_{-25}$, with $n({\rm H})$=60~cm$^{-3}$
and $n({\rm H_2})$=90~cm$^{-3}$. The CN column density towards the star
HD~154368 is 2.7$\times 10^{13}$~cm$^{-2}$ and the line width is
1.2~kms$^{-1}$ \citep{ritchey11}. Interestingly, this source also
shows the second highest excitation temperature $T_{01}$(CN)=2.911$\pm
0.004$~K, which is significantly larger than the weighted mean value
of 2.754~K. Using the physical conditions determined by
\cite{sonnentrucker07}, we have found that an electron density of
0.3~cm$^{-3}$ is necessary to reproduce the measured $T_{01}$ towards
HD~154368. This corresponds to an electron fraction $n(e)/n({\rm
  H_2})\sim 3\times 10^{-3}$, which is too high with respect to the
available carbon. \cite{ritchey11} obtained an even larger value of
0.69~cm$^{-3}$ for HD~154368 and concluded that it probably
corresponds to an upper limit considering the dispersion of 134~mK. In
fact, for this source, \cite{palazzi90} have detected a weak emission
of CN: the strongest hyperfine component $(N, J, F)=(1, 1.5, 2.5)\to
(0, 0.5, 1.5)$ at 113.49~GHz with an antenna temperature $T^*_R=19\pm
5.1$~mK. Figure~\ref{fig:tmb.eps} shows that this value (blue hatched
zone) is reproduced for an electron density of $\sim 3\times
10^{-2}$~cm$^{-3}$, corresponding to an electron fraction $n(e)/n({\rm
  H_2})\sim 3\times 10^{-4}$, as expected if carbon is fully
ionized. In addition, the corresponding excitation temperature is
$T_{01}=2.75$~K, in very good agreement with the weighted mean value
of 2.754~K determined by \cite{ritchey11}.

In summary, our calculations suggest that in the diffuse cloud regions
where CN resides, the electron density lie in the range $n(e)\sim
0.01-0.06$~cm$^{-3}$. This range is significantly smaller than that
derived by \cite{BV1991}, $n(e)\sim 0.02-0.5$~cm$^{-3}$, reflecting
the low (mean) excess temperature $T_{\rm loc}$=29$\pm 3$~mK derived
by \cite{ritchey11}. On the other hand, for individual sources, we
have shown that the dispersion of the optical measurements
($\sim$134~mK) must be taken into account, as recommended by
\cite{ritchey11}. In fact, the weak millimeter emission of CN probably
provides the best accurate measurement of $T_{\rm loc}$, which in turn
yields an accurate determination of $n(e)$ if the kinetic temperature
and hydrogen density is known. Thus, the electron density $n(e)\sim
3\times 10^{-2}$~cm$^{-2}$ derived for HD~154368 might represent the
best indirect measurement of electron density in a diffuse cloud.

\section{Conclusions}

In this work we present a comprehensive set of rates for
fine-structure and hyperfine-structure resolved electron-impact
rotational excitation of the CN radical. Similar rates have previously
been used in an attempt to determine electron densities from shocked
regions of the interstellar medium \citep{jt392,jt477}. Here we
consider the observed temperature excess of CN in diffuse clouds over
the cosmic microwave background.  Assuming this excess is due to
electron and neutral collisions, with electron-impact being
predominant, our calculations suggest that the electron density lies
in the range $n(e)\sim 0.01-0.06$~cm$^{-3}$ for typical physical
conditions present in diffuse clouds. This range of values is
consistent with the known abundance of carbon which is thought to be
the main source of free electrons. We suggest that our methodology
provides a viable means of determining electron densities in the
diffuse interstellar medium.

\section{Acknowledgements}
We thank Fran\c{c}ois Lique for useful discussions.
This work has been supported by STFC through a studentship
to SH and the French CNRS national programme
``Physique et Chimie du Milieu Interstellaire'' (PCMI).

%\begin{section}*{References}
%\References
\bibliographystyle{mn2e}
%\bibliography{journals_phys,rates,jtj,alex}
%\end{section}

\end{document}